# A SERS Investigation of Cyanide Adsorption and Reactivity during the Electrodeposition of Gold, Silver and Copper from Aqueous Cyanocomplexes Solutions

Benedetto Bozzini*, Lucia D'Urzo, Claudio Mele, Vincenzo Romanello

Dipartimento di Ingegneria dell'Innovazione, Università di Lecce

Via Monteroni, I-73100 Lecce (Italy)

* corresponding author: benedetto.bozzini@unile.it

*ABSTRACT*

In this paper we report on the reactivity of adsorbed cyanide deriving from ligand release during metal electrodeposition from cyanocomplex solutions of Au(I), Au(III), Ag(I) and Cu(I) in $H_2O$ and $D_2O$. When $CN^-$ is adsorbed at cathodic potentials in excess of the HER threshold, metal-dependent reactivity can be detected by SERS. Finite surface coverages with adsorbed $CN^-$ at such cathodic potentials can be obtained only if $CN^-$ is delivered directly to the cathode surface as by decomplexing of the cyanocomplexes of the metals undergoing cathodic reduction. In Au(I) and Au(III) baths, Au-$CN^-$ reacts with Au-$H^\bullet$ and is hydrogenated to adsorbed $CH_2$=NH and $CH_3$-$NH_2$. In Ag(I) baths, Ag-$CN^-$ reacts with Ag-$H^\bullet$ giving rise to polycyanogens. No reactivity of Cu-$CN^-$ was found, under otherwise identical conditions. Our conclusions are supported also by dedicated DFT molecular computations.

*KEYWORDS:* hydrogenation, polycyanogen, Raman, DFT, in-situ

## 1. INTRODUCTION

Electrodeposition of Au, Ag, Cu and their alloys is a topic of applied metal electrochemistry with a remarkable commercial background in: electronic, electric, aircraft, medical instrumentation, jewellery and fashion industries. Plating is usually performed in order to protect or decorate items as well as to increase their corrosion or wear resistance, thermal and electric properties or to enhance solderability. Many typical industrial plating processes are carried out with cyanoalkaline solutions containing free $CN^{-}$. Neutral and acidic baths based on Au(I) and Au(III) cyanocomplexes and not containing free $CN^{-}$ incur industrial interest mainly for environmental and personnel safety reasons; some patents and proven processes are available, but their implementation at the production level is impaired chiefly by difficulties in controlling the plating process, as well as by the relevant market structure. At the time of this writing, no industrially relevant plating processes have been implemented making use of neutral Ag and Cu cyanocomplexes, because the commercial standard are free-cyanide alkaline solutions, giving still higher safety concerns. At this stage of the knowledge of these systems, industrial development issues call for fundamental understanding of several interfacial aspects of the electroplating process.[1,2] The understanding of the electrochemical deposition of coinage metals is of both technological and fundamental importance. Extensive efforts have been made towards the understanding of the electrochemical and chemical behaviour involved in the deposition process from cyanocomplexes, as well as on the impact of plating conditions on the functional properties of the electrodeposits. In situ Surface Enhanced Raman Spectroscopy (SERS) is an adequate method for such an approach, in terms of surface sensitivity to adsorption and interfacial reaction phenomena, as well as some degree of quantitative capability.[3,4-11] In the present work, based on in situ SERS, we obtained novel insight into the adsorption and reactivity of $CN^{-}$ at the surface of a growing Au, Ag and Cu cathodes.

## 2. MATERIALS AND METHODS

SERS spectra were recorded using a LabRam microprobe confocal system. A 50x long-working-distance objective was used and the excitation line at 632.8 nm was provided by a 12 mW He-Ne laser. The slit and pinhole were set at 200 and 400 μm, respectively, corresponding to a scattering volume of ~3 pL; Raman spectra were acquired with a 600 grid/mm spectrometer. The recorded Raman intensities are directly proportional to the discharge current of the CCD detector. In situ electrochemical measurements were performed in Ventacon® glass cells with gold, silver and copper disc electrodes (Ø 5 mm) embedded in Teflon holders. The metal surfaces were polished to a mirror finish by a metallographic polishing procedure, which consisted in wet grinding with SiC papers down to 2400 grit number. SERS activity was achieved during the electrodeposition process, via the formation of newly-formed SERS-active metal clusters. The counter electrode was a Pt wire loop of total area ca. 2 $cm^2$, concentric and coplanar with the gold working electrode. An external Ag|AgCl reference electrode was used and all the potentials were reported on this scale. CV measurements were carried out in the cell used for Raman spectroscopy with a scan rate of 10 $mV \cdot s^{-1}$. The electrodeposition baths used are reported in Table 1. All the experiments were run at room temperature (ca. 25°C). Au and Cu electrodeposition experiments were carried out onto Au and Cu electrodes, respectively. Ag electrodeposition experiments were carried out onto both Ag and Au electrodes. The experiments were repeated several times with freshly and independently prepared aliquots of 15 mL of solutions; invariably, reproducibile results were obtained. Ag electrodeposition experiments – yielding the most counterintuitive outcomes - were cross-checked by running repeated and independent measurements also with different Ag and Au cathodes and at two different pH values.

The quantum chemical calculations were performed by the Gaussian 03 package[12] in the framework of the Density Functional Theory (DFT), with the hybrid functional B3LYP and basis set LANL2DZ. In case of the study of polycyclic cyanogens, the PM3 method was used, since semiempiric methods have been proved to be suitable for large molecules and polycondensate ring systems.[13]

According to[14,15] we simulated the adsorption of the cyanide group and its hydrogenation products by attaching one metal atom to the relevant moiety via the carbon atom. We calculated the optimised geometries and the vibrational behaviour of the following structures (where M = Ag, Au, Cu):

1. M-C≡N (unreacted adsorbed cyanide);
2. M-CH=NH (partially hydrogenated adsorbed cyanide)
3. M-$CH_2$-$NH_2$ (fully hydrogenated adsorbed cyanide).

Furthermore, according to[16], the optimised geometries and the vibrational behaviour of two reference structures were studied to take in account the formation of polycyanogens with typical linear structures (LS) and cyclic structures (CS). The vibrational behaviour corresponding to the isotopic substitution - due to the $D_2O$ bath - has been calculated by the "freqcheck" routine.[12] To correct the systematic overestimation of the calculated frequencies, all the results have been rescaled according to the used functional and basis sets, as in[17].

# 3. RESULTS AND DISCUSSION

In order to pinpoint the electrochemical kinetics, we recorded cyclic voltammograms in the potential range investigated by SERS (Figure 1). All the systems exhibit essentially the same principal voltammetric features: (i) the reversible peak $A_1/A_2$; (ii) the partially irreversible peak B and (iii) the Tafel-type growth C. In the Ag system the peak B is not obvious because - owing to the higher electrodeposition overvoltage combined with a relatively low hydrogen overvoltage - it overlaps with the hydrogen evolution c.d.. Feature B is well visible in the Cu system, since the hydrogen overvoltage of Cu is considerably higher. Concerning the Au system, the voltammetric details observed can be explained in terms of the electrodeposition mechanism from Au(I) complexes discussed in[18] for the case of neutral citrate baths and in[19] for cyanoalkaline solutions. Peaks $A_1/A_2$ can be assigned to Au reduction through adsorptive decomplexing of $Au(CN)_2^-$ to $AuCN_{ads}$. The electrochemical reactions B and C correspond to the direct reduction of the cyanocomplexes and to hydrogen evolution, respectively. In the Ag system the $A_2$ oxidation peak appears at potentials more negative than those required for the formation of Ag oxide (ca. 0 V[20]) and can be interpreted with the formation of $Ag(CN)_2^-$.[21,22] Also the Cu voltammogram can be modelled with a two-step reduction process. In fact, reduction can take place from the different kinds of Cu(I) cyanocomplexes that are stable within a given range of $[Cu(I)]/[CN^-]$.[23-25] The progressive decrease of [Cu(I)] in the catholyte with the attending interfacial injection of $CN^-$ as a consequence of the progress of the cathodic process, gives rise to a decrease of the $[Cu(I)]/[CN^-]$ ratio causing the stabilisation of $Cu(CN)_4^{3-}$. The considerably lower electrocatalytic efficiency of Cu is witnessed by the displacement of feature C to higher cathodic values.

We measured in-situ SERS spectra by applying potentiostatic staircases in the potential range -300÷-1650 mV at intervals of 100 mV. The spectra reported are representative of highly reproducible steady-states: (i) transients of spectral patterns and relative intensities were recorded, but only the steady-states will be discussed in this paper; (ii) cathodic-going staircases were reversed in the anodic direction, in order to assess any serial correlation effects, in fact minor hysteretic behaviour was found, that does not affect our conclusions; (iii) statistically independent experiments were run from replicate baths obtained from different batches of chemicals and we repeated all the experiments several times on different days with different cells, cell mountings and electrodes. Furthermore, isotopic shifts were monitored by using $H_2O$ and $D_2O$ as solvents.

3.1 Au electrodeposition from $Au(CN)_2^-$ and $Au(CN)_4^-$ solutions

In Figure 2 we report SERS spectra collected during electrodeposition from the following Au-electrodeposition systems: Au(I)/$H_2O$ (spectra A and B), Au(I)/$D_2O$ (spectrum C) and Au(III)/$H_2O$ (spectrum D). At cathodic polarisations below -1100 mV (e.g. spectrum A) – with both $H_2O$ and $D_2O$ baths -, the only spectral features found relate to intra- and extramolecular vibrations of adsorbed $CN^-$. A broad background band can be noticed, centred at ca. 1000 $cm^{-1}$. The intensity of this broad band varies with the applied potential, as well as with the chemical details for the electrodeposition bath (see also Section 3.2 and 3.3). A detailed description of this phenomenon and its physical interpretation is beyond the scope of the present paper and has been in part addressed in [26], for the case of the Au(I) and Au(III) systems. This band-like background can be due either to a continuum of SERS bands[27] or to adsorbate-induced fluorescence.[28,29] In the present case we are lean to support the latter explanation on the following bases: (i) the fact that the relationship between the degree of surface-enhancement and the background intensity is not a straightforward positive correlation; (ii) from our unpublished DFT computations regarding Au, as well as from literature quantum-chemical results about Ag,[30] $CN^-$ chemisorbed onto metal clusters exhibits electronic transitions that justify fluorescence.

Spectra obtained in the potential range -1100÷-1650 mV with $H_2O$ and $D_2O$ as solvents exhibit new bands (spectra B and C). As the cathodic polarisation is varied in this range, these new bands show potential-independent peak positions accompanied by relative intensity variations with respect to $\nu(CN^-)$ and among themselves. The same behaviour is found with $Au(CN)_4^-$/$H_2O$ solutions (spectrum D). For cathodic polarisations corresponding to the direct discharge of the Au cyanocomplex, the $CN^-$ stretching bands exhibit the linear potential-dependence shown in Figure 3, corresponding to the Stark-tunings listed below, with their 99% confidence intervals and squared linear correlation coefficient $\rho^2$. (i) Au(I)/$H_2O$ 51.2±1.8 $cm^{-1}$ $V^{-1}$ ($\rho^2$=0.994), (ii) Au(I)/$D_2O$ 53.1±2.3 $cm^{-1}$ $V^{-1}$ ($\rho^2$=0.991), (iii) Au(III)/$H_2O$ 46.5 ± 0.47 $cm^{-1}$ $V^{-1}$ ($\rho^2$=1.00). Cases (i) and (iii) match well with our previous results of [26,31]. No statistically significant isotopic effect on the Stark-tuning could be detected, notwithstanding the fact that it could be expected from theoretical considerations.[32] Also extramolecular bands exhibit minor Stark-shifts, coherent with the experimental observations of[18] and with the theory of[33], yielding essentially the same information as the intramolecular one.

On the basis of literature[14,18,34-36] and our DFT computations, we propose the band assignment summarised in Table 2, implying the presence of the following adsorbed species at the growing Au cathode: $CN^-$, $CH_3$-$NH_2$, $CH_2$=NH.

The new bands found during Au electrodeposition at cathodic potentials exceeding -1100 mV and their isotopic shifts can be explained on the basis of the hydrogenation of the $CN^-$ triple bond by $H^\bullet/D^\bullet$ radicals.[37] $CN^-$ hydrogenation would thus be a reaction of $H^\bullet$, alternative to the hydrogen evolution reaction. Our experimental findings support the following reaction path :

$$\mathrm{Au-C\equiv N \xrightarrow{2H^+} Au-CH=NH}$$

$$\mathrm{Au-CH=NH \xrightarrow{2H^+} Au-CH_2-NH_2}$$

<u>3.2 Ag electrodeposition from $Ag(CN)_2^-$ solutions</u>

In Figure 4 we show SERS spectra recorded while electrodepositing Ag from $Ag(CN)_2^-/H_2O$ and $Ag(CN)_2^-/D_2O$ electrolytes. As in the case of Au electrodeposition (see Section 3.1), at cathodic potentials lower than -1100 mV (e.g. spectrum A) both baths with $H_2O$ and $D_2O$ as solvents only bands corresponding to adsorbed $CN^-$ are visible. A less obvious background at low Raman shifts can be noticed.

The spectra measured at cathodic polarisations exceeding -1100 mV using $H_2O$- and $D_2O$-based baths show new bands (spectra B and C), whose intensity grows upon shifting the potential in the cathodic direction at the expence of the $\nu(CN^-)$ peak. These bands correspond to a surface species - of course, within the limits to surface-specificity typical of SERS and Normal Raman Spectroscopy - since spectra measured in the bulk did not show them. It is worth noting that by "bulk spectra" we mean the ones obtained by shifting the confocal volume into the bulk solution enough not to include the surface and by increasing to confocal volume from 3 to 6 pL".

The same behaviour is found at pH 7 and 12. The $CN^-$ stretching bands exhibit a parabolic potential dependence (see Figure 5, for theoretical details on this point, see[32] and references therein contained), the linear ($\alpha$) and parabolic ($\beta$) parameters are listed below, with their 99% confidence intervals and squared linear correlation coefficient $\rho^2$. (i) Ag(I)/$H_2O$ pH 7, $\alpha$: -47.4±5.8 $cm^{-1}$ $V^{-1}$, $\beta$: -34.6±2.7 $cm^{-1}$ $V^{-2}$ ($\rho^2$=0.999); (ii) Ag(I)/$D_2O$ pH 7, $\alpha$: -43.9±4.9 $cm^{-1}$ $V^{-1}$, $\beta$: -33.1±2.4 $cm^{-1}$ $V^{-2}$ ($\rho^2$=0.998); (iii) Ag(I)/$H_2O$ pH 12, $\alpha$: -29.9±5.7 $cm^{-1}$ $V^{-1}$, $\beta$: -26.7±2.8 $cm^{-1}$ $V^{-2}$ ($\rho^2$=0.998). Case (iii) matches well with our previous results of[16,21]. As far as isotopic effects and extramolecular vibrations are concerned, the same comments as in Section 3.1.

On the basis of literature[16,38] and original DFT computations, we propose the band assignment listed in Table 3. The fact that many bands related to $CN^-$ reaction products are not affected by isotopic shift can be rationalised in terms of the formation of polycyanogens. We carried out DFT computations of typical linear (LS) and polycyclic condensed (CS) polymers proposed by[16], whose optimised structures are displayed in Figure 6. Our DFT simulations of the CS have been carried out with 7 rings because computations of the relevant vibrational properties - performed with structures containing a number of rings from 1 to 15 - showed an essentially asymptotic behaviour after 7 rings.

The new bands found during Ag electrodeposition at potentials lower than -1100 mV show a reasonable matching with structures LS and CS of[16] and can be explained with the formation of linear and cyclic polycyanogens of different molecular weights and degrees of branching. Polycyanide formation processes have received limited attention in the literature and are typically based on cyanogen polymerisation, no reports of relevant electrochemical processes are known to the authors. A possible cathodic route to polycyanogen formation can be the following. At cathodic potential high enough for the reduction of $H^+$, the following reaction can be assumed to occur at a Ag surface covered by adsorbed cyanide:

$$\mathrm{Ag\text{-}C \equiv N \xrightarrow{H^+/e^-} Ag\text{-}H + CN^{\bullet}}$$

The possibility of desorbing $CN^{\bullet}$ from metal-CN species – though under very special electronic-structure conditions - has been proved in.[39,40] $CN^{\bullet}$ is liable to attack $CN^{-}$ giving rise to polymerisation by radical attack, according to the conceptual scheme:

$$\mathrm{Ag\text{-}C \equiv N \xrightarrow{CN^{\bullet}} Ag\text{-}C^{\bullet} = N\text{-}C \equiv N \xrightarrow{2n\,CN^{\bullet}} Ag\text{-}\left[C\left(C \equiv N\right) = N\right]_n\text{-}C \equiv N}$$

<u>3.3 Cu electrodeposition from $Cu(CN)_2^-$ solutions</u>

In Figure 7 we show SERS spectra recorded while electrodepositing Cu in a $Cu(CN)_2^-/H_2O$ bath. At variance with the cases of Au and Ag electrodeposition (see Sections 3.1 and 3.2) only bands corresponding to adsorbed $CN^-$ are visible both at low (e.g. -550 mV, spectrum A) and high (e.g. -1650 mV, spectrum B and -1850 mV, spectrum C) cathodic polarisations. We measured SERS spectra up to -1850 mV - beyond which level SERS data could not be recorded owing to excessive lowering of the signal-to-noise ratio resulting from scattering by hydrogen bubbles -, but we could not find any qualitative spectral change. Of course, the $Cu(I)/CN^-$ system is less noble than the Ag and Au ones, and the hydrogen overvoltage for Cu is higher than for Ag and Au, nevertheless - in the investigated potential range - both Cu(I) and $H^+$ reduction are active, but unable to give rise to reactivity of $CN^{-}_{ads}$ observable by SERS. A strong background at low Raman shifts can be noticed at low cathodic overvoltages. At higher cathodic potentials the ν(CN) peak is multiple, coherently with.[41]

3.4 General discussion

The kind of reactivity of $CN^-$ highlighted in the present research has not been observed in the past, to the best of the authors' knowledge. The fact that these particular types of reactivity of the C≡N bond during cathodic processes in aqueous media during Au and Ag electrodeposition has not been pinpointed previously, is probably due to the following reasons. (i) In conventional $CN^-$ adsorption work with KCN and NaCN solutions without concomitant reduction of a cyanocomplex, in the high cathodic potential range - where the hydrogen evolution reaction takes place - $CN^-$ is desorbed. Moreover, in thin-layer spectroelectrochemical work, such as IR or SFG, field confinement occurs and electrode area where the hydrogen evolution reaction takes place is expelled from the analysed region.[23] To the contrary, in the case of electrodeposition from cyanocomplexes, transiently adsorbed $CN^-$ is present also at very high cathodic potentials, as a consequence of the ongoing electroreduction process, implying the release of $CN^-$ to the surface. (ii) For the first time SERS work during electrodeposition has been carried out with baths containing – apart from the metal complexes – exclusively non-specifically-adsorbing supporting electrolyte instead of the conventional organics, that are liable to offer alternative routes for the reactivity of the adsorbed $H^{\bullet}$ radicals.

In the foregoing discussion, we highlighted profoundly different outcomes of $CN^-$ adsorbed onto Au, Ag and Cu at cathodic polarisations up to -1650 mV. For the special case of Cu, the potential has been shifted to higher cathodic values (i.e. up to -1850 mV): it is not the purpose of this paper to be exhaustive in putting forward a definitive explanation of this sensitivity to the electrodeposited metals, under otherwise identical bath chemistry and electrochemical polarisation, but we reckon that a simple tentative explanation can be the different electrocatalytic activity of these metals with respect to the hydrogen evolution reaction. This electrocatalytic activity is well-known to rank in the order: Au>Ag>>Cu, e.g. according to the "Volcano-plot" representation, expressing a positive correlation between M-$H_{ads}$ bond strength, or $\Delta H_{ads}$, on one side and hydrogen overvoltage and exchange current density on the other one.[42,43] The hydrogen-evolution mechanism on Au and Ag electrodes in acidic solutions has been reported to be the Tafel-Volmer route,[44] to the best of the authors' knowledge no such information is available regarding Cu, but – from the literature Tafel slopes[45] – one can conclude that the mechanism is the same also for this metal. These literature data are coherent with our voltammetric work for the system currently investigated (see Figure 1). Correspondingly: (i) hydrogenation of adsorbed $CN^-$ seems to be favoured on Au; (ii) attack of $CN^-$ by $H^\bullet$, leading to the formation of $CN^\bullet$ active towards polymerisation was found to take place on Ag and (iii) no reactivity of $CN^-$ could be pinpointed in the hydrogen-evolution range of potentials onto Cu. Alternatively - in order to explain the differences in reactivity towards $CN^-_{ads}$ observed among the three metals we studied - a photoelectrochemical route can be envisaged. In fact: (i) the rate of $H^\bullet$ formation at an Au electrode can be enhanced by photogeneration[46] and (ii) surface-catalysed photopolymerisation of polycyanogen might occur, that would render the process viable also with a wavelength such as 632 nm, longer than the range reported to be effective in[19]. Further possible explanations could also include differences in the degree of surface-enhancement among the metals investigated, as well as differences in the respective dielectric functions, brought about by roughening during electrodeposition. On the basis of the qualitative differences found in our data among the studied metals, we note that such catalytic effects lead to differences in reaction pathways rather than simply on reaction rates.

## 4. CONCLUSIONS

In this paper we carried out an in-situ SERS investigation of Au, Ag and Cu electrodeposition processes from Au(I), Au(III), Ag(I) and Cu(I) cyanocomplex solutions with $H_2O$ and $D_2O$ as solvents. Our chief results are:

(i) At low cathodic polarisations, adsorbed $CN^-$ is observed in all the systems investigated.

(ii) At high cathodic potentials – in the case of Au and Ag electrodeposition, adsorbed $CN^-$ tends to react with hydrogen radicals - formed in the concurrent proton reduction process – giving rise to reactions products that can be explained with hydrogenation in the case of Au and polymerisation in that of Ag. No such reactivity is found with Cu, under otherwise identical electrochemical conditions, as well as under extreme cathodic polarisations.

SERS spectra suggest that the hydrogenation products obtained in Au(I)/$H_2O$, Au(I)/$D_2O$ and Au(III)/$H_2O$ baths are identical in structure and – on the basis of the spectroscopy literature and original DFT calculations - can be identified with $CH_2$=NH and $CH_3$-$NH_2$.

At high cathodic polarisations, neutral and alkaline Ag(I) baths prepared with $H_2O$ and $D_2O$, give rise to polymers that – again on the basis of the limited cognate literature and dedicated DTF work – can be explained with a mixture of linear and cyclic polycondensed polycyanogens.

To the best of the authors' knowledge this type of cathodic reactivity has not been highlighted in the past, this finding is liable to have a remarkable bearing on: (i) optimisation and control of many industrial processes – that, in fact, exhibit operational instabilities and product inconsistencies -, (ii) environmental issues related to cyanide management and (iii) opens the possibility of new synthetic routes, based on the combination of the in-situ formation of $H^\bullet$ radicals and the special catalytic activity of newly-formed metal clusters.

| | Electrodeposition bath | pH |
|---|---|---|
| (i) | $KAu(CN)_2$ 10 mM, $NaClO_4$ 0.1 M, $H_2O$ | 7 |
| (ii) | $KAu(CN)_2$ 10 mM, $NaClO_4$ 0.1 M, $D_2O$ | 7 |
| (iii) | $KAu(CN)_4$ 10 mM, $NaClO_4$ 0.1 M, $H_2O$ | 7 |
| (iv) | $KAg(CN)_2$ 10 mM, $NaClO_4$ 0.1 M, $H_2O$ | 7 |
| (v) | $KAg(CN)_2$ 10 mM, $NaClO_4$ 0.1 M, $D_2O$ | 7 |
| (vi) | $KAg(CN)_2$ 10 mM, $NaClO_4$ 0.1 M, $H_2O$ | 12 by KOH |
| (vii) | $KCu(CN)_2$ 10 mM, $NaClO_4$ 0.1 M, $H_2O$ | 7 |

Table 1 – Electrodeposition baths used in this work

| Mode | (A) cm$^{-1}$ | (B)Au(I)/$H_2O$ cm$^{-1}$ | (C) Au(I)/$D_2O$ cm$^{-1}$ | (D) Au(III)/$H_2O$ cm$^{-1}$ |
|---|---|---|---|---|
| Au-$CH_2NH_2$ bending | | 270<br>DFT=287 | 249<br>DFT=253 | 273 |
| Au-C≡N Bending | 298 [14,18,36] | | | |
| Au-C of Au-C≡N stretching | 370 [14,18,36] | | | |
| Au-$CH_2NH_2$ stretching | | 468<br>DFT=456 | 421<br>DFT=431 | 469 |
| -$NH_2$ "out of plane" bending | | 640<br>DFT=628 | 530<br>DFT=538 | 640 |
| Au-CH=NH CH wagging | | 905<br>DFT=957 | 760<br>DFT=758 | 910 |
| Au-CH=NH CH wagging | | 1005<br>DFT=986 | 864<br>DFT=898 | 1010 |
| C-N stretching | | 1200<br>DFT=1180 | 990<br>DFT=957 | 1210 |
| $CH_2$ scissoring | | 1440<br>DFT=1430 | 1110<br>DFT=1084 | 1440 |
| Au-$CH_2$-$NH_2$ NH scissoring | | 1600<br>DFT=1626 | 1180<br>DFT=1211 | 1600 |
| $CH_2$ symm. stretching | | 2850<br>DFT=2850 | | |
| $CH_2$ asymm. stretching | | 2900<br>DFT=2930 | 2165<br>DFT=2171 | |
| hydoxyl stretching | | Ca. 3200[34,35] | Ca. 2400[46] | |

Table 2 – Band assignments for SERS spectra measured in the $Au(CN)_2^-/H_2O$, $Au(CN)_2^-/D_2O$ and $Au(CN)_4^-/H_2O$, baths (Figure 2). The simulated frequency values have been calculated in the framework of Density Function Theory with B3LYP functional and Lan2DZ basis set and rescaled according to[17].

| **Mode** | **(A) cm⁻¹** | **(B) Ag(I)/$H_2O$ cm⁻¹** | **(C) Ag(I)/$D_2O$ cm⁻¹** |
|---|---|---|---|
| Ag-C≡N Bending[14] | 240 | | |
| Ag-C of Ag-C≡N stretching[14] | 313 | | |
| LS#13 combination bending | | 325<br>DFT(LS)=324 | 325 |
| LS#14 C-C≡N bending | | 366<br>DFT(LS)=369 | 366 |
| LS#16 C-C≡N bending, CS#25 condensed ring stretching | | 436<br>DFT(LS)=461<br>DFT(CS)=427 | 436 |
| CS#31 condensed ring out of plane symmetric bending | | 501<br>DFT(CS)=507 | 501 |
| LS#17 combination of C-C≡N bending, C-N stretching and C=N stretching | | 555<br>DFT(LS)=544 | 555 |
| LS#19 skeleton bending | | 609<br>DFT(LS)=585 | 609 |
| CS#47 condensed ring out of plane asymmetric bending | | 782<br>DFT(CS)=802 | 782 |
| LS#24 N-C=N bending, CS#55 condensed ring asym stretching | | 837<br>DFT(LS)=839<br>DFT(CS)=844 | 837 |
| CS#61 condensed ring quadrant stretching | | 1005<br>DFT(CS)=1005 | 1005 |
| CS#62 condensed asymm. ring stretching | | 1080<br>DFT(CS)=1094 | 1080 |
| LS#28 C-C-N stretching | | 1132<br>DFT(LS)=1147 | 1132 |
| CS#65 aromatic asymm. CC stretching | | 1203<br>DFT(CS)=1200 | 1203 |
| LS#29 N-C=N bending, CS#66 asymm. quadrant stretching | | 1264<br>DFT(LS)=1255<br>DFT(CS)=1266 | 1264 |
| CS#70 asymm. C-N stretching | | 1345<br>DFT(CS)=1362 | 1345 |
| CS#74 asymm. C-C stretching | | 1414<br>DFT(CS)=1418 | 1414 |
| CS#76 symm. C-C stretching | | 1450<br>DFT(CS)=1443 | 1450 |
| CS#78 condensed symm. C-C stretching | | 1480<br>DFT(CS)=1483 | 1480 |
| Hydroxyl bending[38] | | 1584<br>DFT=1500 | 1185<br>DFT=1150 |
| LS#32 asymm. C-C=N stretching, CS#81 asymm. C=N stretching | | 1635<br>DFT(LS)=1642<br>DFT(CS)=1662 | 1655 |
| CH symm/asymm stretching[34] | | 2817/2874 | 2407/2503 |
| hydroxyl stretching[38] | | 3221[34,35] | 2670[47] |
| NH stretching | | 3453[34] | 2890<br>DFT=2800 |

Table 3 – Band assignments for SERS spectra measured in the $Ag(CN)_2^{-}/H_2O$ and $Ag(CN)_2^{-}/D_2O$ (Figure 4). Structures LS (linear) and CS (cyclic) are shown in Figure 6, mode codes correspond to those adopted in[12]. The simulated frequency values have been calculated in the framework of Density Function Theory with B3LYP functional and Lan2DZ basis set and rescaled according to.[17] In the case of cyclic structures the PM3 method has been used.

## FIGURE CAPTIONS

Figure 1 - Cyclic voltammograms for the following systems: (i) Au electrode in contact with a solution of composition $KAu(CN)_2$ 10 mM, $NaClO_4$ 100 mM, pH 7; (ii) Ag electrode in contact with a solution of composition $KAg(CN)_2$ 10 mM, $NaClO_4$ 100 mM, pH 7; (iii) Cu electrode in contact with a solution of composition $KCu(CN)_2$ 10 mM, $NaClO_4$ 100 mM, pH 7.

Figure 2 – SERS spectra recorded during Au electrodeposition: (A) $Au(CN)_2^-/H_2O$ bath, -950 mV; (B) $Au(CN)_2^-/H_2O$ bath, -1650 mV; (C) $Au(CN)_2^-/D_2O$ bath, -1650 mV; (D) $Au(CN)_4^-/H_2O$ bath, -1650 mV. Band assignments reported in Table 2.

Figure 3 - Potential dependence of the $CN^-$ stretching SERS band for the Au electrodeposition systems: squares Au(I)-$H_2O$, circles Au(I)-$D_2O$, triangles Au(III)-$H_2O$.

Figure 4 – SERS spectra recorded during Ag electrodeposition: (A) $Ag(CN)_2^-/H_2O$ bath, -600 mV; (B) $Ag(CN)_2^-/H_2O$ bath, -1300 mV; (C) $Ag(CN)_2^-/D_2O$ bath, -1300 mV. Band assignments reported in Table 3.

Figure 5 - Potential dependence of the $CN^-$ stretching SERS band for the Ag electrodeposition systems: squares Ag(I)-$H_2O$ pH 7, circles Ag(I)-$D_2O$ pH 7, triangles Ag(I)-$H_2O$ pH 12.

Figure 6 - The optimised geometry of polycyanogens with linear structure (LS) and cyclic structure (CS), as simulated by DFT/B3LYP - LANL2DZ.

Figure 7 – SERS spectra recorded during Cu electrodeposition: $Cu(CN)_2^-/H_2O$ bath (A) -550 mV, (B) -1650 mV and (C) -1850 mV.

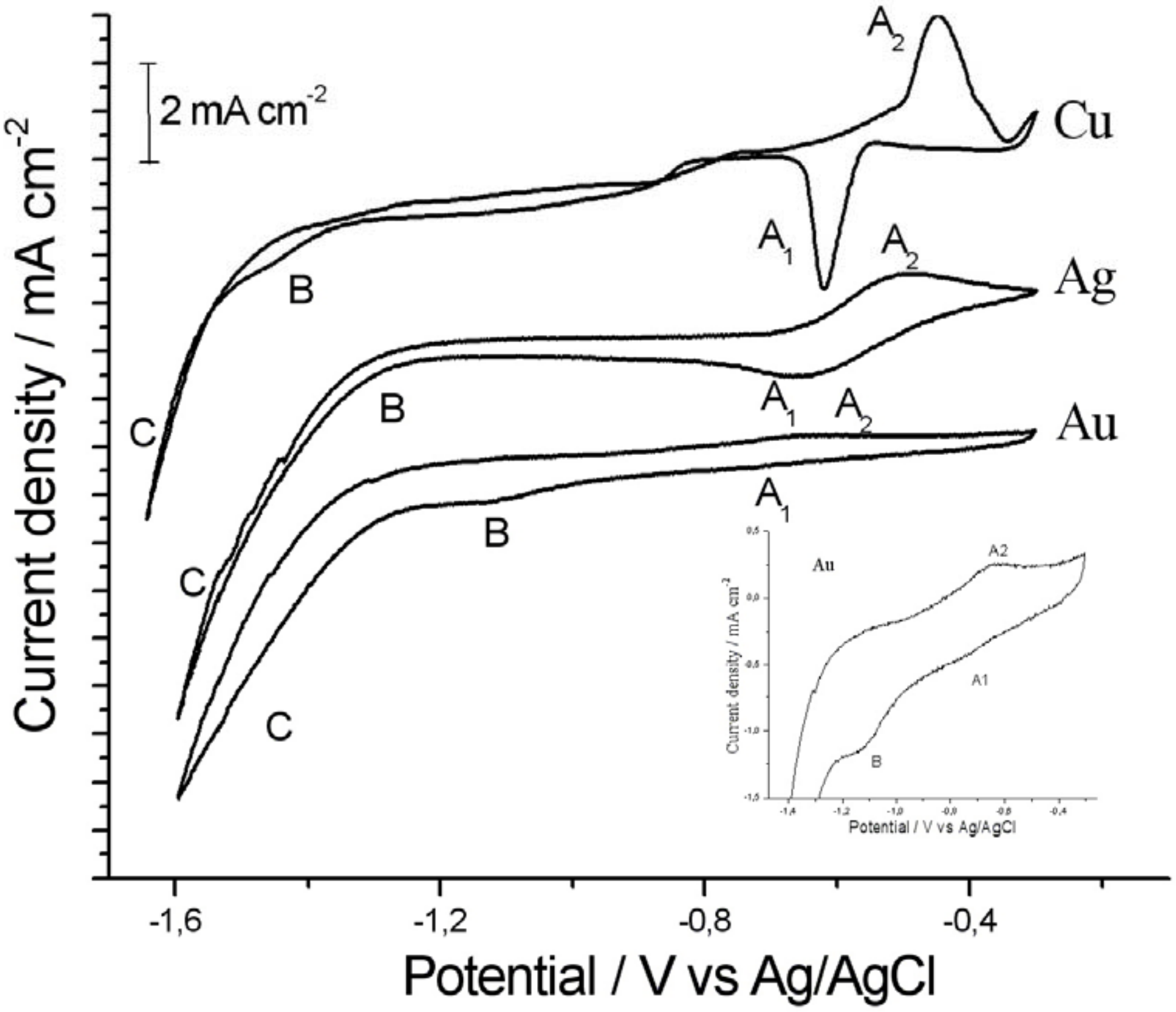

Figure 1

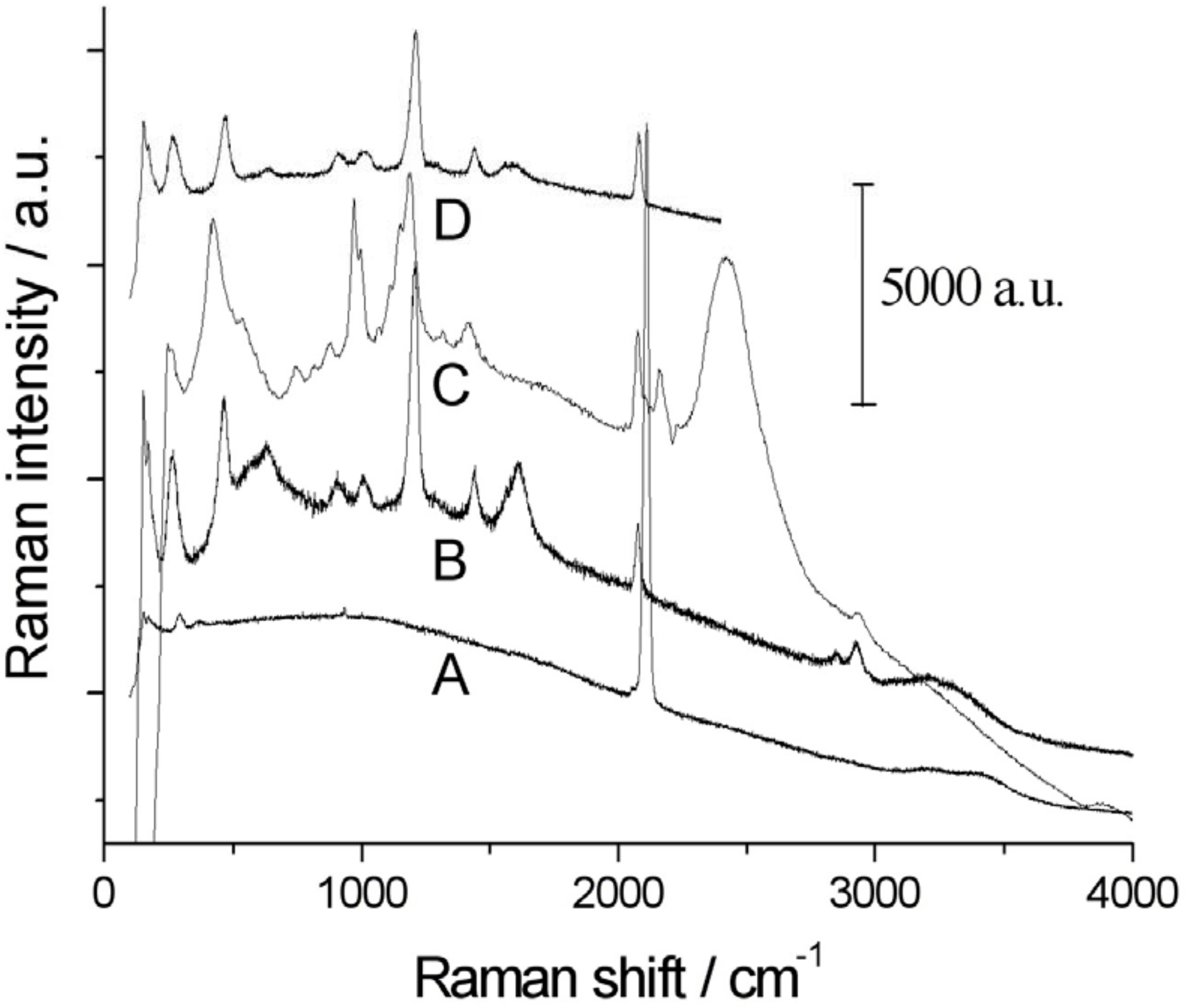

Figure 2

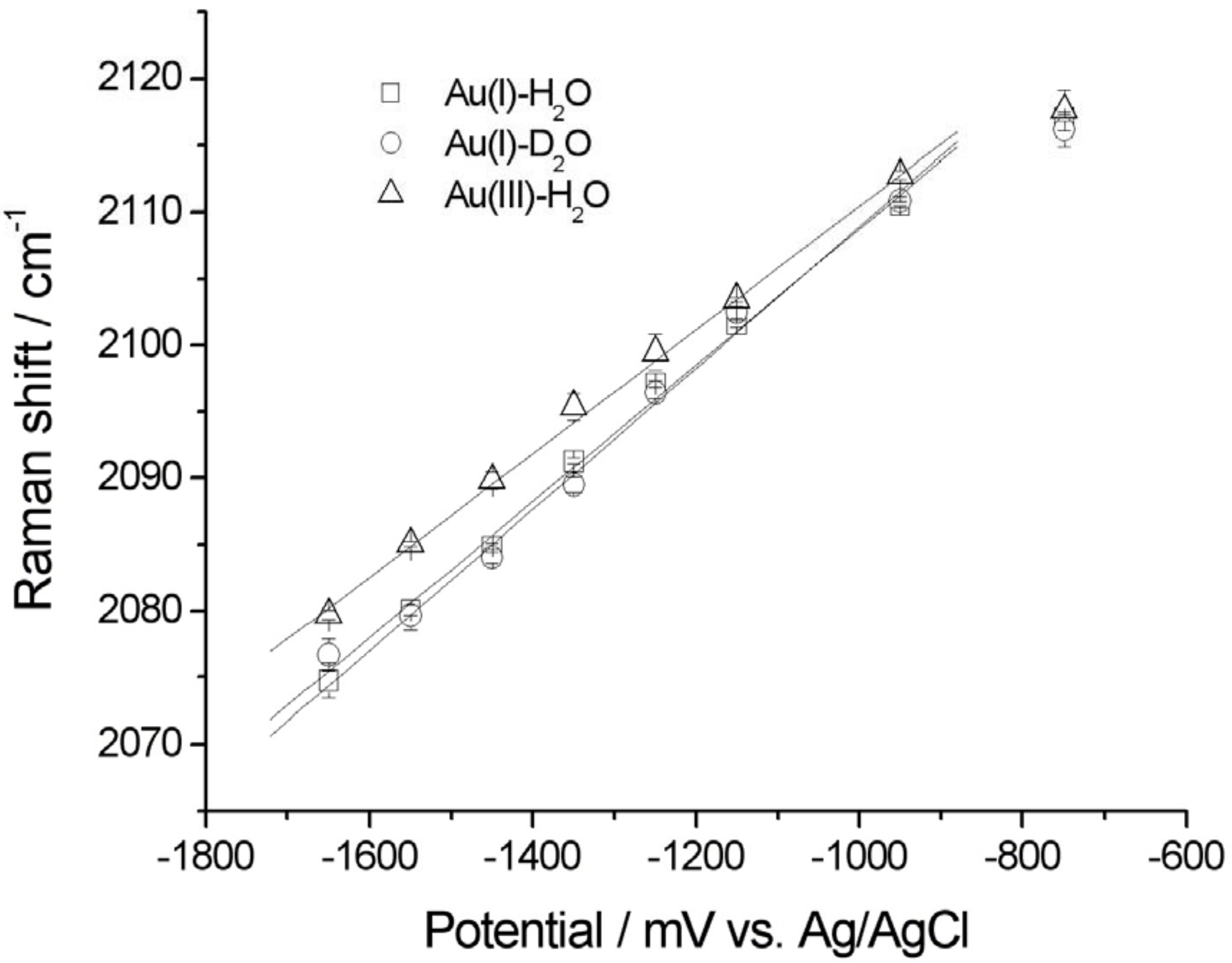

Figure 3

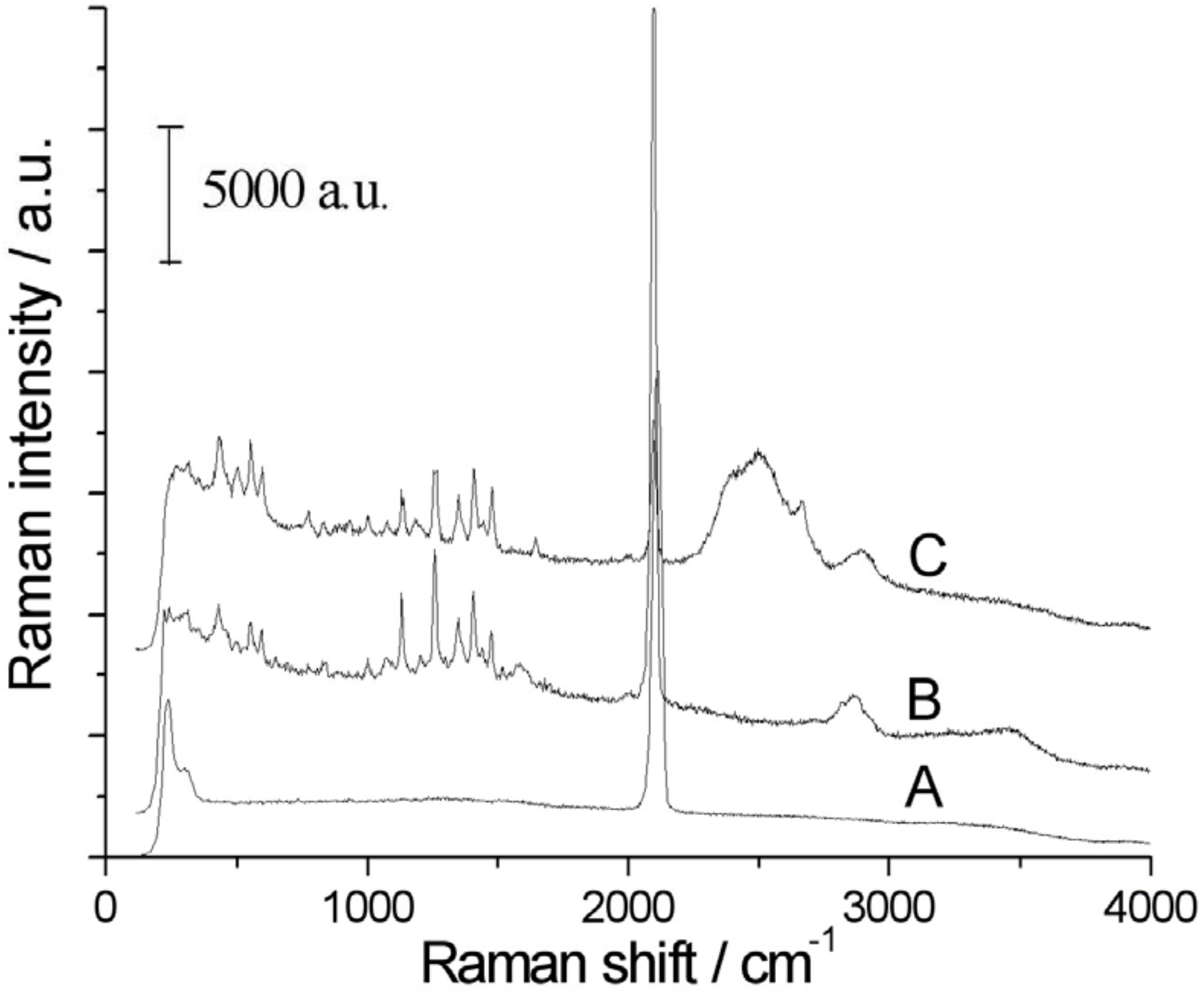

Figure 4

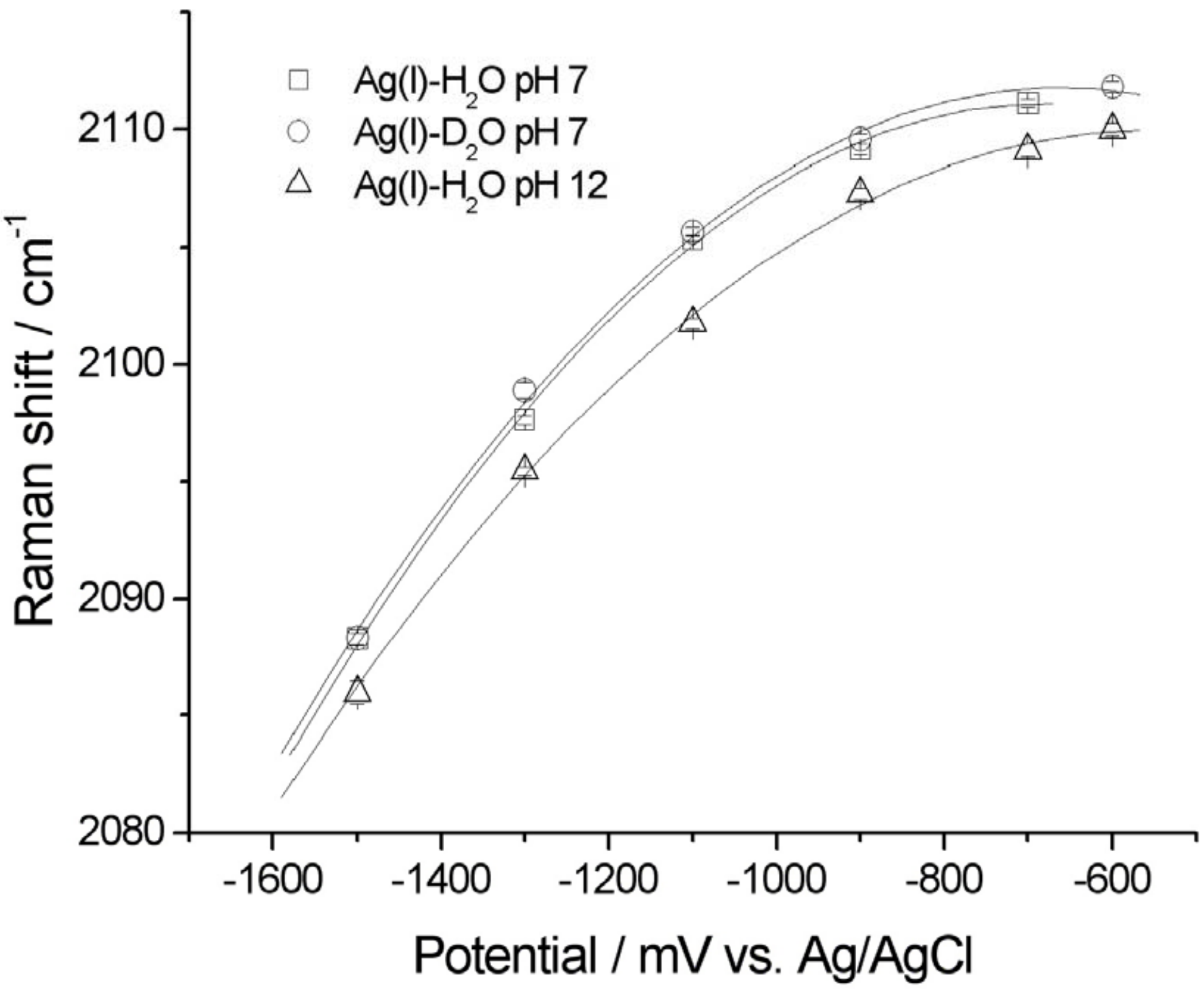

Figure 5

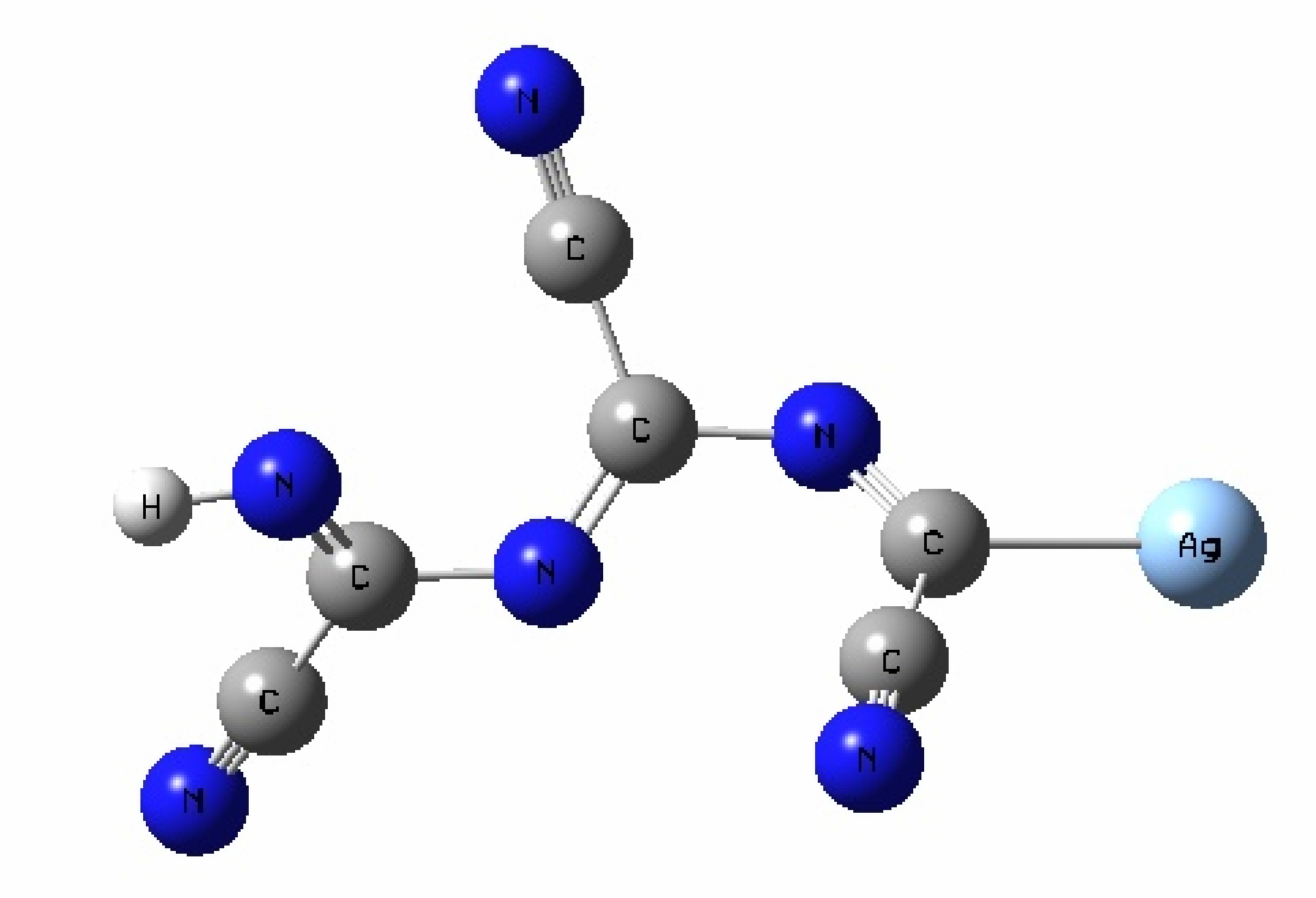

**LS**

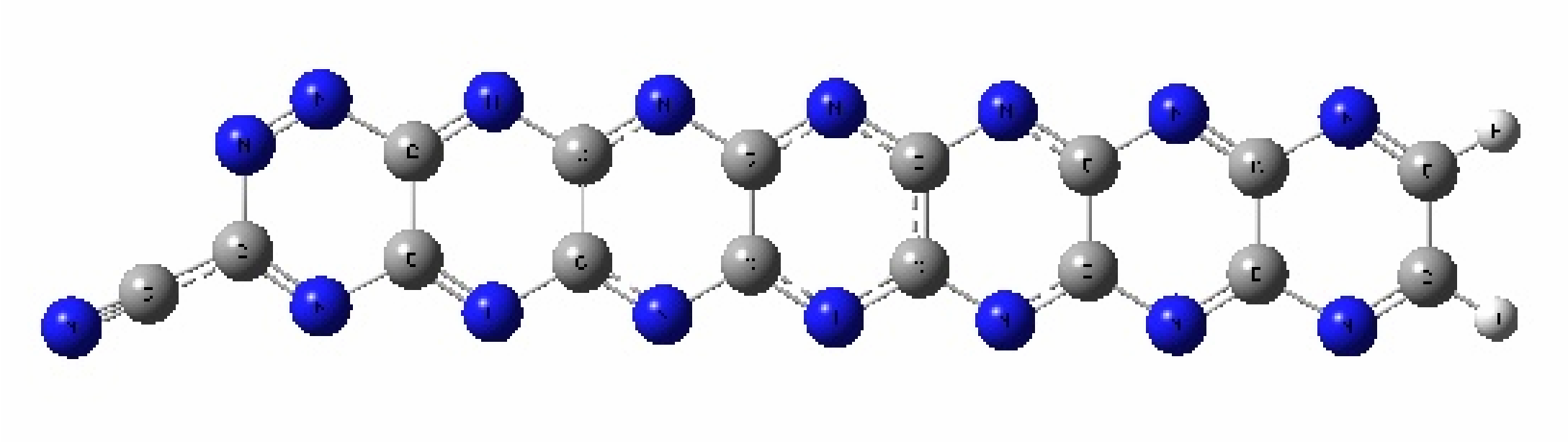

**CS**

Figure 6

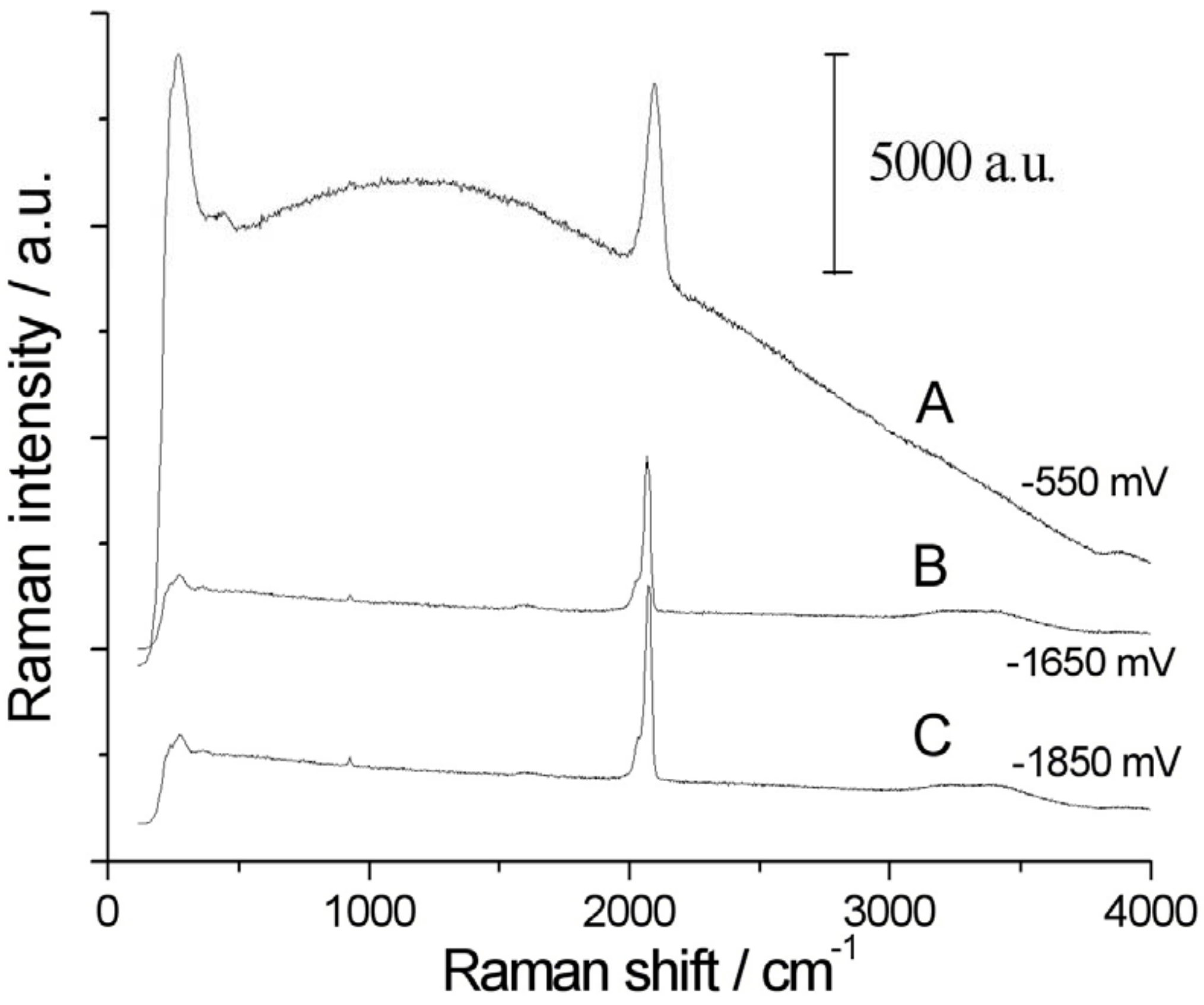

Figure 7